\documentclass[aps,prd,twocolumn,showpacs,superscriptaddress,nofootinbib,preprintnumbers]{revtex4-2}

\usepackage{style}

\begin{document}


\title{Boiling After the Dust Settles: \\ Constraining First-Order Phase Transitions During Dark Energy Domination}

\author{Seth Koren}
\author{Yuhsin Tsai}
\affiliation{Department of Physics and Astronomy, University of Notre Dame, South Bend, IN 46556}
\author{Runqing Wang}
\affiliation{Department of Physics and Astronomy, University of Notre Dame, South Bend, IN 46556}
\affiliation{Department of Physics and Center for Field Theory and Particle Physics, Fudan University, 2005
Song Hu Road, Shanghai 200438, China}

\date{\today}

\begin{abstract}
A first-order phase transition could occur in the late universe when vacuum energy begins dominating the energy density ($z \lesssim 0.3$) and convert some latent heat into other forms such as invisible radiation. 
This generic possibility also has concrete motivation in particle physics models which invoke a multitude of vacua to address theoretical puzzles.
The na\"{i}ve constraint on such an event comes from measurements of the Hubble expansion rate, but this can only probe transitions involving $\mathcal{O}(10)\%$ of the dark energy.
In this work, we show that significantly tighter constraints appear when accounting for phase transition fluctuations affecting CMB photon propagation anisotropically, akin to the integrated Sachs-Wolfe effect. 
For instance, if a completed phase transition has $\beta/H_\star\lesssim 25$, current CMB data limits the associated vacuum energy released to less than $1\%$ of the dark energy.
A transition to negative vacuum energy (quasi-anti-de Sitter) is allowed only for $\beta/H_\star \gtrsim 300$. For $\beta/H_\star \lesssim 500$, the universe will not crunch for at least $14$ Gyr.
\end{abstract}
\bigskip
\maketitle

\section{Introduction}\label{sec:intro}
Cosmology provides a powerful window into the dark sector, where new particles—including dark matter—can interact strongly with each other while remaining nearly invisible to the Standard Model (SM) particles. Decades of direct, indirect, and collider searches suggest that any interaction with SM particles must be extremely weak, if present at all. So far, gravity remains our most reliable tool for probing the dark sector. Fortunately, precision cosmological observations—sensitive to both the evolution and fluctuations of the universe—allow us to study gravitational fluctuations induced by dark sector dynamics across a wide range of scales and epochs. In particular, they open the door to exploring dramatic events such as first-order phase transitions (FOPTs) that may occur entirely within the dark sector.

FOPTs have long been studied as efficient sources of gravitational waves (GWs). Transitions occurring at SM photon temperatures around $100$~GeV to $1$~TeV can produce signals detectable by future GW detectors like LISA and BBO~\cite{Kosowsky:1992rz,Kosowsky:1991ua,Kosowsky:1992vn,Kamionkowski:1993fg, Caprini:2015zlo,Caprini:2019egz,Caldwell:2022qsj,LISA:2022kgy, LISACosmologyWorkingGroup:2022jok}, with anisotropies that can reflect non-trivial inflationary or reheating dynamics~\cite{Geller:2018mwu,Kumar:2021ffi,LISACosmologyWorkingGroup:2022kbp,Bodas:2022urf,Cui:2023dlo}. FOPTs at MeV to GeV scales have been suggested as explanations for the stochastic GW background observed by Pulsar Timing Arrays~\cite{NANOGrav:2020bcs, NANOGrav:2023gor, EPTA:2021crs, EPTA:2023fyk, EPTA:2023xxk, Goncharov:2021oub, Reardon:2023gzh, Xu:2023wog,Bringmann:2023opz, Addazi:2023jvg, Ghosh:2023aum,Salvio:2023ynn, Winkler:2024olr,Banik:2025fnc}. In the eV range, they have been proposed to help resolve the Hubble tension~\cite{Riess:2021jrx,Schoneberg:2021qvd,Niedermann:2019olb,Niedermann:2020dwg,Niedermann:2023ssr,Garny:2024ums,Chatrchyan:2024xjj}, and may also produce observable CMB B-mode signals in future measurements~\cite{Greene:2024xgq}. Due to strong constraints from Big Bang Nucleosynthesis (BBN) and the CMB, FOPTs below the $10$~MeV scale are generally required to occur entirely within a dark sector to avoid excessive reheating of SM particles~\cite{Bai:2021ibt}. 

During FOPTs, bubble nucleation is stochastic and so occurs anisotropically even in the absence of primordial adiabatic perturbations. The resulting bubble wall collisions, sound waves, and magnetohydrodynamic motions of the plasma generate both scalar and tensor perturbations. While tensor modes, especially their role in producing GW, have been extensively studied~\cite{Caprini:2007xq,Jinno:2016vai,Jinno:2017fby}, scalar perturbations, including curvature and isocurvature modes, remain much less explored. Recent studies~\cite{Liu:2022lvz,Elor:2023xbz,Lewicki:2024ghw,Buckley:2024nen,Zou:2025sow,Franciolini:2025ztf} have begun filling this gap, showing that precise measurements of the CMB anisotropies, energy spectrum, and matter power spectrum already provide tight constraints on FOPTs occurring between $1$~eV and $10$~keV temperature. Additional constraints arise from sub-horizon density inhomogeneities generated during the PT, such as plasma sound waves~\cite{Ramberg:2022irf} or curvature perturbations sourced by gravitational waves from bubble collisions~\cite{Geller}.

In this work, we take a step further and explore extremely late-time FOPTs at redshifts $z \lesssim 0.3$, within the dark energy domination of the last few billion years. We derive upper bounds on the change in vacuum energy from a dark sector phase transition using constraints from CMB anisotropy measurements. 
We consider the effect of inhomogeneous redshifting of CMB photons due to variations in the time they enter true vacuum regions, and show that this can impose much stronger constraints than those from the modification of cosmic expansion. 

To isolate this universal effect of late-time FOPTs, we model the latent heat as being converted into free-streaming radiation. This means that the density perturbations are quickly damped out after the PT completes, so the fluctuations imprinted on the CMB photons arise \textit{only} from the time variation in the `first-encounter-surface' of the PT (see Fig. \ref{fig.sketch}). 
In future work we will consider the particular case where latent heat is converted into slowly-moving fluids, which allows density perturbations to persist and leads to even more stringent constraints.

As to why such late FOPTs are well motivated to study and constrain, we recall the earlier periods in the conventional model of the early universe when vacuum energy nears domination and note that each time, this portends new vacuum dynamics which turn this latent heat into dynamical degrees of freedom. 
This may occur by the rolling of an inflationary scalar field to its minimum; or by a crossover as at the QCD phase transition; or by a violent FOPT as in proposals for electroweak baryogenesis.
Quite recently in the history of the cosmos we have seen vacuum energy re-emerge as a dominant component, so it seems reasonable to ask whether this too may signal a recent or future phase transition. 

Further beyond the SM, one large class of solutions to naturalness problems of particle physics uses large numbers of vacua and dynamical evolution between them. This strategy was famously invoked for the cosmological constant problem \cite{Abbott:1984qf,Brown:1987dd,Brown:1988kg,Duncan:1989ug,Bousso:2000xa,Feng:2000if,Dvali:2001sm,Kaloper:2022jpv,Kaloper:2022yiw,Kaloper:2023kua,Kaloper:2023xfl,Kaloper:2025goq}, and some mechanisms for the hierarchy problem \cite{Dvali:2003br,Dvali:2004tma,Graham:2015cka,Kaplan:2015fuy,Choi:2015fiu,Ibanez:2015fcv,Gupta:2015uea,Hook:2016mqo} and the strong CP problem \cite{Banks:1991mb,Dvali:2005zk,Kaloper:2017fsa,Kaloper:2025upu,Kaloper:2025wgn} work similarly. Generically, within such scenarios it is possible to have a further vacuum transition at a recent time. This includes transitions to \textit{negative} vacuum energy, which lead eventually to crunching universes, on which we focus below in Sec. \ref{sec:ads}.

Of course neither of these general motivations tells us precisely when such a phase transition should occur or exactly its nature.
Here we consider the more-violent case of a first-order phase transition, and its observational signatures when we can test it: if it has already happened. 
We emphasize that our purpose is not to propose specific models, but rather to analyze the signals and constrain generic models in which such behavior would occur.
In comparison to the quintessence case of late vacuum dynamics \cite{Fujii:1982ms,Ford:1987de,Ratra:1987rm,Wetterich:1987fm}, this has seen far less study. 

Finally, there is some empirical motivation from recent late-time observations of DESI that hint at non-trivial dynamics of dark energy~\cite{DESI:2025zgx}. As for now the best-fitting model of these dynamics is far from clear, but it would be interesting to study models incorporating a FOPT. 

\section{Fluctuations of the phase transition time}
A FOPT occurs through the nucleation and expansion of bubbles, the interiors of which are in the true vacuum. This process is inherently stochastic, so different regions transition at slightly different times. The bubble nucleation rate per unit time and volume is described by~\cite{HOGAN1983172, PhysRevD.45.3415,Hindmarsh:2015qta}:  
\begin{equation}\label{eq.Gamma}
\Gamma = \Gamma_0 e^{-S(t)} \approx \Gamma_0 e^{-S(t_f)}e^{\beta(t-t_f)}\,,
\end{equation}
where $S$ is the bounce action for creating a critical bubble, and $\beta \approx -{\rm d} S(t_f)/{\rm d} t$. The timescale $t_f$ serves as a reference time marking the PT's progression, defined as when an $e^{-1}$ fraction of space remains in the false vacuum~\cite{ATHRON2024104094}.
When $\beta \gg H_\star \equiv H(t_f)$, where $H(z)$ is the Hubble scale, the transition completes quickly, finishing in a small fraction of the Hubble time near $t_f$. 

To estimate the redshift fluctuations of CMB photons entering true vacuum regions, we begin by evaluating the fluctuations in the PT time at different spatial points. We define the local transition time at position $\vec{x}$ as $t_c(\vec{x})$, and quantify its fluctuations using the two-point function $H_\star^2 \langle \delta t_c(\vec{x}) \delta t_c(\vec{y}) \rangle$, where $\delta t_c(\vec{x}) = t_c(\vec{x}) - \bar{t}_c$ and $\bar{t}_c$ is the average transition time. The corresponding dimensionless power spectrum is derived in~\cite{Elor:2023xbz} via its Fourier transform:
\begin{equation}\label{eq.Pdt}
\mathcal{P}_{\delta t}(k) = \frac{k^3}{2\pi^2} \left(\frac{H_\star}{\beta}\right)^2 \int \mathrm{d}^3 r~ e^{i \vec{k} \cdot \vec{r}} \beta^2 \langle \delta t_c(\vec{x}) \delta t_c(\vec{y}) \rangle,
\end{equation}
with $\vec{r} = \vec{x} - \vec{y}$. For perturbation modes with comoving wavelengths larger than the average comoving bubble separation $k^{-1}\gg d_b\approx(8\pi)^{1/3} (1+\bar{z}_{\rm pt})v_w\beta^{-1}$~\cite{PhysRevD.45.3415,Hindmarsh:2015qta}, where $\bar{z}_{\rm pt}$ is the redshift at which the PT completes, the power spectrum primarily reflects the standard deviation of the PT completion time averaged over a volume $\sim k^{-3}$. This leads to the dimensionless power spectrum~\cite{Elor:2023xbz}:  
\begin{equation}\label{eq:Pdt}  
{\cal P}_{\delta t}(k) \approx 3(8\pi) v_w^3 \left(\frac{(1+\bar{z}_{\rm pt})k}{H_\star}\right)^3 \left(\frac{H_\star}{\beta}\right)^5\,,  
\end{equation}
assuming a bubble wall velocity $v_w$. For wavelengths shorter than the average bubble separation $k^{-1}\ll d_b$, the power spectrum is proportional to ${\cal P}_{\delta t}(k)\propto [(1+\bar{z}_{\rm pt})k/H_\star]^{-3}(H_\star/\beta)^{-1}$. 
In Appendix~\ref{app.A} we show the full $\delta t_c$ spectrum we use in our calculations, adapted from~\cite{Elor:2023xbz}.

\section{CMB Redshift Fluctuations from Phase Transition}\label{sec:Hubble}
\begin{figure}
\includegraphics[width=8cm]{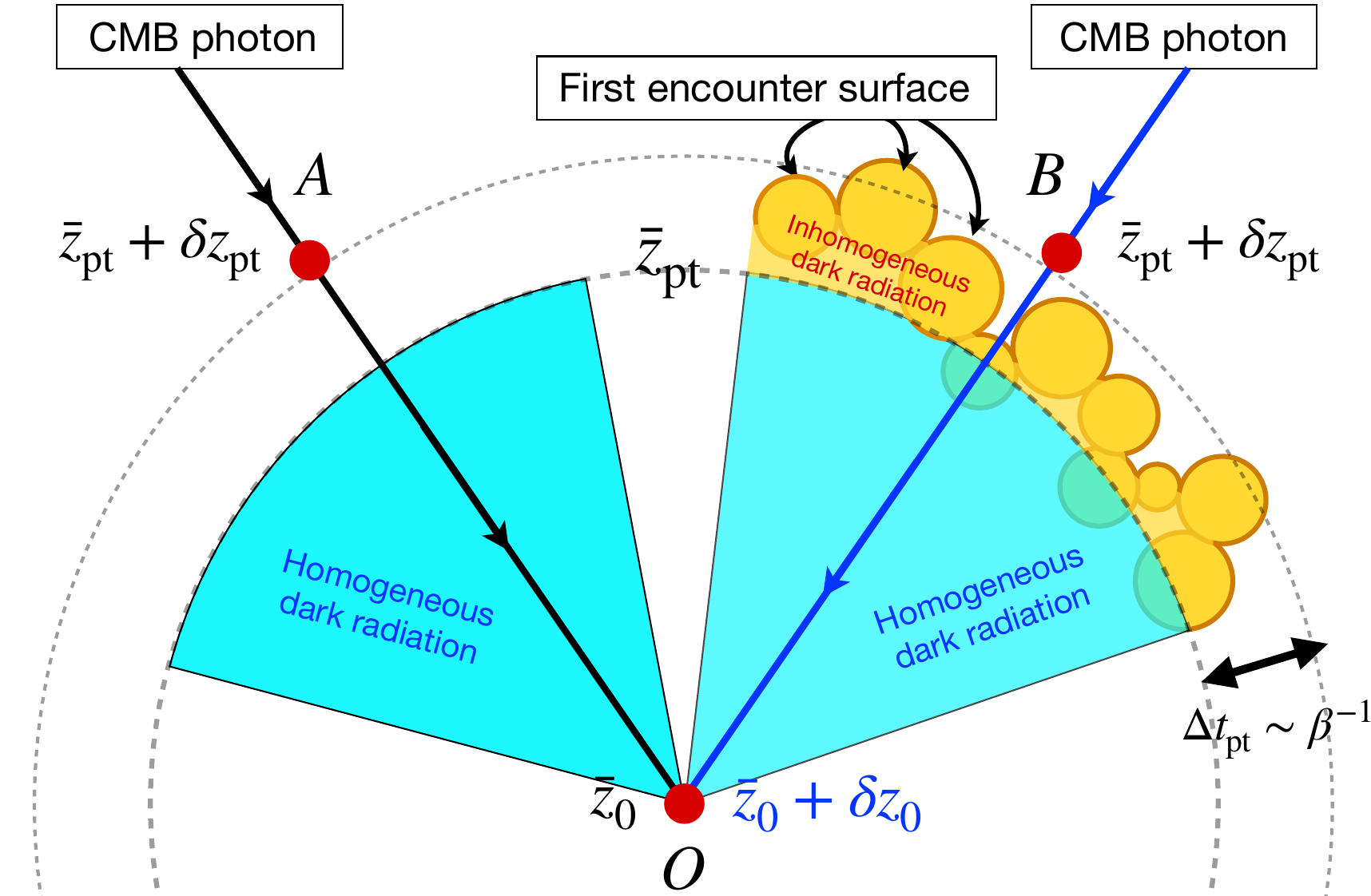}
\caption{Sketch of the origin of CMB fluctuations from a late-time FOPT, comparing photon trajectories before (left) and after (right) accounting for stochastic bubble nucleation. See text for details.}
\label{fig.sketch}
\end{figure}
We consider a FOPT in a dark energy-dominated universe, with the transition to the true vacuum completing at redshift $\bar z_{\rm pt}$. 
Across the PT a fraction $r>0$ of the dark energy density is released, and we note $r > 1$ means the vacuum energy becomes negative which is physically allowed. To isolate the minimal signatures we consider a scenario where the latent heat is converted into free-streaming, massless dark radiation though in more general scenarios richer physics may occur. We model the PT as leading to a patchwork of regions with slightly different Hubble parameters due to the variance in the local PT time, leading to differential local Hubble expansion as follows: 
\begin{eqnarray}\label{eq.Hz}
&H_{\rm pt}&(z,z_{\rm pt},r)
\\
&\equiv& H_0\sqrt{\Omega_\Lambda(1-r)+r\,\Omega_\Lambda\left(\frac{1+z}{1+z_{\rm pt}}\right)^4+\Omega_m(1+z)^{3}}\,,\nonumber
\end{eqnarray}
where we use the local $z_{\rm pt}$ (without a bar) to describe the spatially-varying dark radiation density, discussed further below. Because the PT considered here only affects the CMB spectra at $\ell \lesssim 50$, while most $\Lambda$CDM parameter constraints are determined by higher $\ell$ modes, we adopt the Planck~2018 best-fit parameters \cite{Planck:2018vyg}: $H_0 = 68{\rm \ km \ s^{-1}Mpc^{-1}}$, $\Omega_\Lambda = 0.69$, and $\Omega_m = 0.31$ in our analysis. Rather than performing a full fit of the $\Lambda$CDM+PT model to cosmological data, we estimate upper bounds on $r$ from CMB anisotropy using these fixed parameters. As we will show, even when saturating the anisotropy bounds, the change in the universe's average redshift due to the PT can be well below $1\%$, resulting in shifts in $\Lambda$CDM energy densities smaller than the current precision. As the CMB fluctuations we consider come from random bubble distributions of PT, which are uncorrelated to the adiabatic perturbations, we turn off the adiabatic perturbations in the following discussion and consider the total CMB power spectrum to be the sum of PT and $\Lambda$CDM results. 

We focus on PTs with $\beta/H_\star\approx 10\text{–}500$, 
which implies that a PT finishing at $\bar{z}_{\rm pt}$ started at nearly the same redshift. A CMB photon traveling along the line of sight will enter the true vacuum region at $\bar{z}_{\rm pt}+\delta{z}_{\rm pt}(\theta,\phi)$, where the redshift at photon entry varies due to stochasticity. We note that as $\bar{z}_{\rm pt}$ is defined at the redshift at which the PT finishes, $\delta z_{\rm pt} \geq 0$. Since the PT quickly takes over the space, the photon will then always be in the true vacuum region. 
As Hubble is decreasing inside, the photons can be redshifted more or less before reaching us, depending on $\delta{z}_{\rm pt}(\theta,\phi)$, which defines a ``first-encounter-surface" (Fig.~\ref{fig.sketch}) that is similar to the last scattering surface of the CMB photon during recombination. 

The first-encounter-surface determines when along a CMB photon's trajectory it sees the background density redshifting, and so the amount of propagation in the redshifting background for a given photon fluctuates by order $\delta z_{\rm pt}$. The stochasticity in the phase transition time also induces spatial variations in the dark sector energy density, which in turn could modify the local Hubble rate at all later times, imprinting additional fluctuations in the photon spectrum. However, since we consider free-streaming dark radiation, the energy density becomes homogenized on the scale of the bubble size, which is where the dominant CMB anisotropy signal arises. 
For this reason, we take a conservative, simplifying assumption by only considering perturbations due to the inhomogeneous redshifting of dark radiation within $[\bar{z}_{\rm pt}+\delta z_{\rm pt}, \bar{z}_{\rm pt}]$ (yellow region in Fig.~\ref{fig.sketch}), which is comparable to the PT duration.
Beyond this time interval, we treat the energy density as homogeneous, and no further photon fluctuations are generated.\footnote{If the PT reheats into slow-moving fluids such as dark matter, the lack of free-streaming damping could preserve larger energy density fluctuations at low redshift, enhancing the CMB anisotropy. The perturbations we study are thus the minimal signatures, and we leave these cases for future work.}

\begin{figure}
    \centering
\includegraphics[width=1\linewidth]{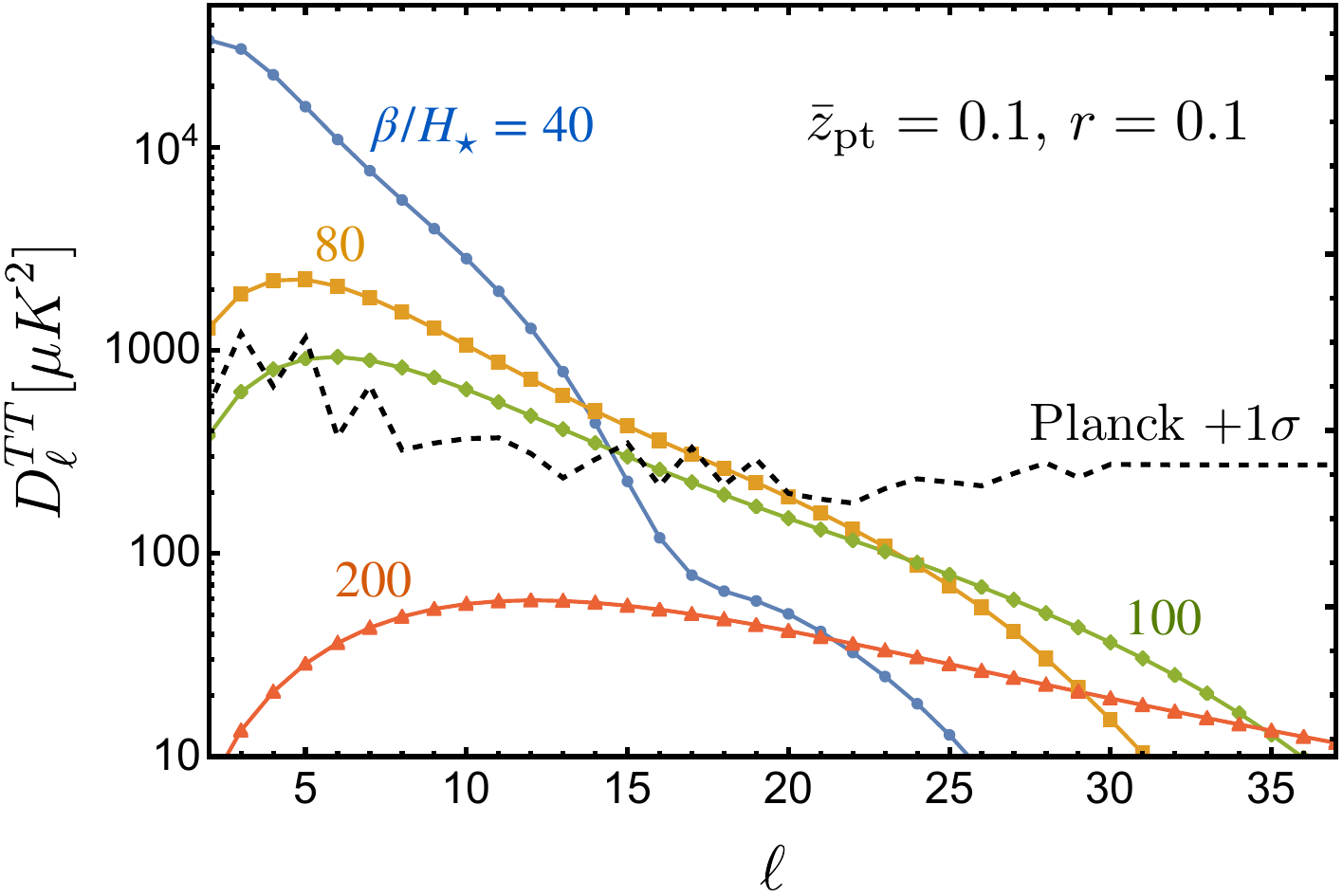}
    \caption{CMB temperature power spectrum from a PT at $\bar{z}_{\rm pt} = 0.1$ with $r = 0.1$. Discrete $\ell$ points are connected for visualization. The $+1\sigma$ error bar from Planck 2018~\cite{planckarchiv} is shown for comparison.}
    \label{fig:DlTT}
\end{figure}

We define the comoving distance 
\begin{equation}
\chi_{z_{\rm pt},r}(z_i,z_f)=\int^{z_i}_{z_f}\frac{{\rm d}z}{(1+z)H_{{\rm pt}}(z,z_{\rm pt},r)}\,.
\end{equation}
In the no-PT case, $r=0$, and we drop the $z_{\rm pt}$ dependence to write $\chi_{z_{\rm pt},r=0}\equiv\chi_0(z_i,z_f)$.\

The first-encounter-surface is anisotropic and located at $\bar{z}_{\rm pt} + \delta z_{\rm pt}(\theta,\phi)$ (line $BO$ in Fig.~\ref{fig.sketch}). The comoving distance from this entry point to us is:
\begin{equation}\label{eq.chieq}
\chi_{BO}\Big|_{\delta z_0}^{\bar{z}_{\rm pt}+\delta z_{\rm pt}}=\chi_{\bar z_{\rm pt},r}(\bar{z}_{\rm pt},\delta z_0)+\chi_{\bar{z}_{\rm pt}+\delta z_{\rm pt},r}(\bar{z}_{\rm pt}+\delta z_{\rm pt},\bar{z}_{\rm pt})\,,
\end{equation}
where the final redshift $\delta z_0$ also depends on $(\theta,\phi)$. 
To determine this fluctuation, we can compare to the case where the PT occurs uniformly at $\bar z_{\rm pt}$ without fluctuations. There are no fluctuations of the final redshift in this case, and $\delta z_0=0$ (line $AO$ in Fig.~\ref{fig.sketch}). The comoving distance a photon travels from redshift $\bar z_{\rm pt} + \delta z_{\rm pt}(\theta,\phi)$ to today is given by
\begin{eqnarray}\label{eq.chiAO}
\chi_{AO}\Big|_{0}^{\bar{z}_{\rm pt}+\delta z_{\rm pt}}&=&\chi_{\bar z_{\rm pt},r}(\bar{z}_{\rm pt},0)+\chi_0(\bar{z}_{\rm pt}+\delta z_{\rm pt},\bar{z}_{\rm pt})\,,\label{eq.chiAO}
\end{eqnarray}

Since the phase transition only affects the expansion rate of the universe, the total comoving distance from the same initial redshift to us must remain the same. This gives the condition:
\begin{equation}\label{eq.chieq}
\chi_{AO}\Big|_{0}^{\bar{z}_{\rm pt}+\delta z_{\rm pt}}=\chi_{BO}\Big|_{\delta z_0}^{\bar{z}_{\rm pt}+\delta z_{\rm pt}}\,.
\end{equation}
Solving this equality gives the final redshift fluctuation $\delta z_0$ as a function of the PT-time fluctuation $\delta z_{\rm pt}$. For small values $\bar z_{\rm pt},\delta z_0,\delta z_{\rm pt}\ll 1$, the solution is well-approximated by:
\begin{equation}\label{eq.dz0}
\delta z_0\approx \delta z_{\rm pt}^2\left[\frac{r\,\Omega_\Lambda}{(1+\bar{z}_{\rm pt})\left[\Omega_\Lambda+\Omega_{\rm m}(1+\bar{z}_{\rm pt})^3\right]^{3/2}}\right]\,.
\end{equation}
This expression agrees with the numerical solution at the $\mathcal{O}(1)\%$ level for the parameters we consider. The fluctuation scales with the PT energy $r$ and with $\delta z_{\rm pt}^2$, where one power of $\delta z_{\rm pt}$ arises from its effect on photon propagation, and the other from having a significant redshift of dark radiation energy in the first $\delta z_{\rm pt}$ window. For $\delta z_{\rm pt}>0$, the CMB photon enters the true-vacuum region earlier, and the Hubble expansion is slower at a later redshift due to a smaller total energy density compared to the uniform-PT case. This requires the expansion to end earlier with $\delta z_0>0$ to keep the same comoving distance. 

We estimate the redshift power spectrum $\mathcal{P}_{\delta z_{\rm pt}}$ using the power spectrum of the PT time at which each point transitions to the true vacuum 
\begin{equation}\label{eq.Pdzen}
\mathcal{P}_{\delta z_{\rm pt}}(k)\approx(1+\bar{z}_{\rm pt})^2{\cal P}_{\delta t}(k)\,.
\end{equation}
Combining this with Eq.~(\ref{eq.dz0}), we compute the two-point function $\langle \delta z_{\rm pt}^2(\theta_1,\phi_1)\delta z_{\rm pt}^2(\theta_2,\phi_2)\rangle$ and obtain the power spectrum of the induced photon redshift:
\begin{equation}\label{eq.Pdz0}
\mathcal{P}_{\delta z_{0}}(k)\approx r^2\,\Omega_\Lambda^2[\Omega_\Lambda+\Omega_{\rm m}(1+\bar{z}_{\rm pt})^3]^{-3}\,{\cal I}_{\delta t}(k)\,,
\end{equation}
where
\begin{equation} \label{eq.Idt}
{\cal I}_{\delta t}(k)\equiv k^3\int_{k_{\rm min}}^{k_{\rm max}}\frac{{\rm d}r}{r}\mathcal{P}_{\delta t}(r)\int_{-1}^1{\rm d}\mu \,s^{-3}\mathcal{P}_{\delta t}(s)\,,
\end{equation}
and $s=\sqrt{k^2+r^2-2kr\mu}$. In the numerical integral, we set $k_{\rm min}$ ($k_{\rm max}$) to be 10 times smaller (larger) than the peak mode of $\mathcal{P}_{\delta t}$, which provides a good approximation. Further details of the derivation of Eqs.~(\ref{eq.dz0}) and (\ref{eq.Pdz0}) are in the Appendix.

The anisotropy we obtain corresponds to a late-time integrated Sachs–Wolfe effect~\cite{Sachs:1967er}, sourced by metric perturbations from the PT around $\bar{z}_{\rm pt}$. The redshift perturbation $\delta z_0$ alters the CMB photon temperature, and its contribution to the CMB power spectrum is estimated by projecting $\mathcal{P}_{\delta z_0}(k)$ onto the first-encounter surface using the standard method
\begin{equation}
D_\ell^{\rm TT, pt}= 2\ell(\ell+1) \int_{k_{\rm min}}^{k_{\rm max}} \frac{{\rm d}k}{k}\mathcal{P}_{\delta z_0}(k) j_\ell^2(k\Delta\tau)\,.
\end{equation}
We use $\Delta\tau=\chi_0(\bar{z}_{\rm pt}, 0)$ as the no-PT case, since the allowed PT only introduces a negligible ($<1\%$) correction to the comoving distance. Since the PT signal we consider mainly peaks at $\ell < 20$, we omit the reionization suppression factor in the calculation. Including  $e^{-2\tau}\approx0.89$~\cite{Planck:2018vyg} for signal peaks at higher $\ell$ would only change the bound on $r$ at the percent level. We show some examples of $D_\ell^{\rm TT, pt}$ in Fig.~\ref{fig:DlTT}. 
\begin{figure}
    \centering    \includegraphics[height=1.0\linewidth]{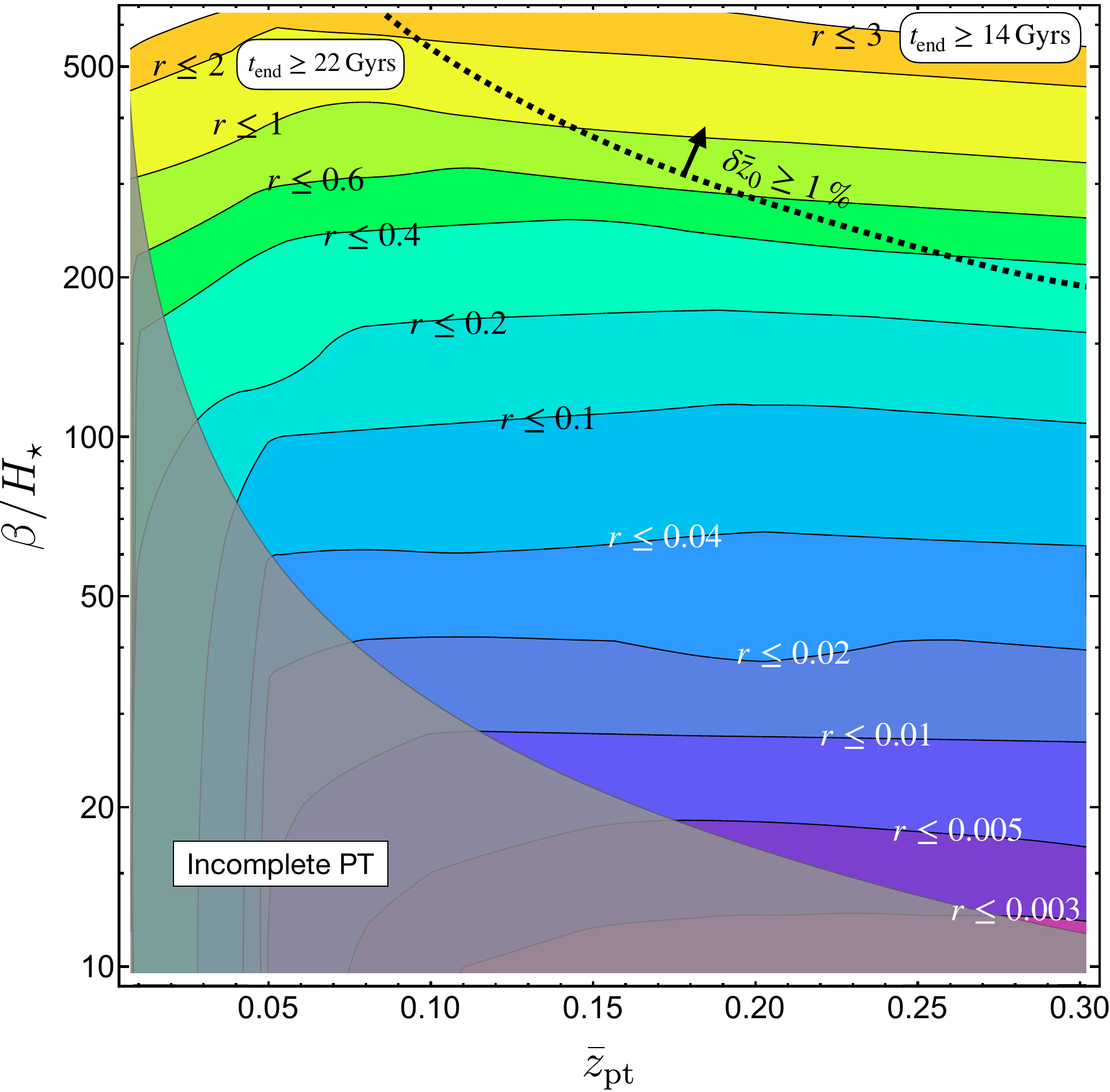}
    \caption{Upper bounds on a FOPT converting a fraction $r$ of dark energy into dark radiation, assuming wall velocity $v_w=1$. The dashed curve indicates where the present-day redshift is shifted by $1\%$ relative to the no-PT, implying a $\approx1\%$ rescaling of $\Lambda$CDM energy densities at earlier times if $z_0=0$ is fixed. The gray region indicates where the distance to $\bar{z}_{\rm pt}$ falls within the average bubble size, where our anisotropy estimate does not apply. For $r>1$, the square boxes show the minimal crunching time from Eq.~(\ref{eq.FE1}).}
    \label{fig:chisquared}
\end{figure}

We estimate the bounds on $r$ using a $\chi^2$-analysis
\begin{equation}
\chi^2 = \sum_{\ell=\ell_p-1}^{\ell_p+1} \frac{\left[D_\ell^{\rm TT,pt}(r,\beta/H_\star,z_{\rm pt})\right]^2}{\sigma_\ell^2}\,,
\end{equation}
where $\sigma_\ell$ are the  1$\sigma$ error bars taken directly from the Planck~2018 archive~\cite{planckarchiv}. The uncertainties include both foreground and cosmic variance. $\ell_p$ is the peak location of $D_\ell^{\rm TT,pt}$. We use three $\ell$-bins for the $\chi^2$ calculation, and require $\chi^2\leq5.99$ to set a $2\sigma$ upper bound on $r$ with given $(\beta/H_\star,\bar{z}_{\rm pt})$. Including more $\ell$-bins only mildly affects the bound. While this method is not fully accurate—since Planck error bars are highly correlated and depend on the assumed baseline cosmology—it still gives a useful estimate of the allowed size of $r$ without significantly modifying the observed $D_\ell^{\rm TT}$.

Fig. \ref{fig:chisquared} shows the upper bounds on $r$, assuming $v_w=1$. A useful analytic order-of-magnitude approximation is 
\begin{equation}\label{eq.rapprox}
r \lesssim 10^{-5}\left(\frac{\beta}{H_{\star}}\right)^{2}\,.
\end{equation}
Approximating the power spectrum with just its peak value around $\delta t_c \sim \beta^{-1}$, we get $\mathcal{P}_{\delta t}\sim (\beta/H_\star)^{-2}$, and Eqs.~(\ref{eq.Pdz0}, \ref{eq.Idt})  give $\mathcal{P}_{\delta z_0}\sim r^2(\beta/H_\star)^{-4}$. The CMB bound is roughly $\mathcal{P}_{\delta z_0}\lesssim A_s \approx 2\times10^{-9}$, where $A_s$ refers to the primordial scalar perturbation, leading to Eq. (\ref{eq.rapprox}). 

The $r$ bounds also become weaker at lower $\bar{z}_{\rm pt}$, as the peak of $\mathcal{P}_{\delta z_0}(k)$ shifts to such low $k$-modes that it no longer contributes significantly to the power spectrum with $\ell\geq2$. 
Note that a naive estimate of the PT bound by requiring a $\lesssim 10\%$ suppression in the Hubble rate from Eq.~(\ref{eq.Hz}) at $z \approx 0.2$, as motivated by the $h(z)/h^{\Lambda{\rm CDM}}$ constraint from the DESI BAO analysis~\cite{DESI:2025fii}, allows a PT at $\bar{z}_{\rm pt} = 0.3$ with $r \approx 0.65$. The CMB anisotropy bounds we obtained are significantly stronger for $\beta/H_\star\lesssim 200$.

Since our estimate assumes a completed PT, the average comoving bubble separation must be larger than the comoving distance from redshift $\bar{z}_{\rm pt}$ to today:
\begin{equation}
d_p\approx(8\pi)^{1/3}v_w\beta^{-1}(1+\bar{z}_{\rm pt})>\Delta\tau.
\end{equation}
Although an incomplete PT will still affect CMB photon propagation and lead to strong constraints on $r$, we leave a detailed analysis of the incomplete PT (grey) regime for future work.

We have so far assumed the existence of a PT and calculated the resulting CMB anisotropies from fluctuations in the transition redshift, $\delta z_{\rm pt}$. Compared to standard $\Lambda$CDM with no PT, the change in $H_{\rm pt}$ also induces a homogeneous redshifting today, $\delta \bar{z}_0$, which we can determine by equating
\begin{equation}\label{eq.avgchi}
\chi_{\bar{z}_{\rm pt}, r}(\bar{z}_{\rm pt}, \delta\bar{z}_0) = \chi_{0}(\bar{z}_{\rm pt},0)\,.
\end{equation}
After rescaling today’s redshift back to zero, the $\Lambda$CDM energy densities at earlier times could be modified at the level of $\delta\bar{z}_0$. For reference, Fig.~\ref{fig:chisquared}  shows a dashed curve for $\delta\bar{z}_0 = 1\%$ when saturating the $r$ bounds from the anisotropy estimate. This overall redshift shift is comparable to uncertainties in current $\Lambda$CDM density measurements, and the region above the curve may be further constrained by fitting $\Lambda$CDM energy densities to the full CMB data.

\section{the end times} \label{sec:ads}
For large $\beta/H_\star$, FOPTs with $r > 1$ become allowed. Here the released latent heat exceeds the dark energy density $\Lambda_0 = \rho_c \Omega_\Lambda$, where $\rho_c = (2.5\,\text{meV})^4$~\cite{ParticleDataGroup:2024cfk} is the critical density, leading the universe to \emph{decelerate} and eventually crunch. This regime may be motivated from a UV perspective, as there is evidence from string theory that true vacua have $\Lambda_0 < 0$, and any positive $\Lambda_0$ must be metastable (see  e.g.~\cite{Obied:2018sgi} and references therein). 
A slowly-rolling scalar whose minimum is eventually found at $V< 0$ is one way such scenarios might be made consistent with our universe---but a late-time phase transition could also satisfy these conjectures.

The second Friedmann equation after the PT is
\begin{eqnarray}
    \frac{\ddot{a}}{a} = -H_0^2\Omega_\Lambda \left[r\left(\frac{1+z}{1+\bar z_{\rm pt}}\right)^4 - (1-r)+\frac{\Omega_m}{2\Omega_\Lambda}(1+z)^{3}\right]\,.\nonumber\\
    \label{eq.FE1}
\end{eqnarray}
For $r>1$ the right-hand side is strictly negative, so the expansion begins decelerating. 

The continued expansion of the universe further dilutes the radiation and matter densities until the universe switches from expanding to contracting when $H(z_{\rm max})^2 = 0$. Neglecting matter density ($\Omega_\Lambda \to 1$), this occurs at
\begin{equation}
\frac{1+\bar z_{\rm pt}}{1 + z_{\rm max}}=\left(\frac{r}{r-1}\right)^{1/4}.
\end{equation}
Following this point, the scale factor decreases, causing the energy density to grow and leading to increasingly rapid contraction. The universe then collapses at a time
\begin{equation}\label{eq.tend}
t_{\rm end}\approx H_0^{-1} \sqrt{\frac{12}{r-1}}\text{ArcCos}\left[ \sqrt{\frac{1}{2}+\frac{\sqrt{2}}{4} \sqrt{1-\frac{1}{\sqrt{r}}}}\right].
\end{equation}

In the yellow and orange regions of Fig.~\ref{fig:chisquared} where $r>1$ is allowed, we show the lower bounds on $t_{\rm end}$ corresponding to the upper bounds $r\leq2$ and $ \leq 3$, obtained by numerically solving Eq.~(\ref{eq.FE1}). The anisotropy bound ensures that, for a PT with $\beta/H_\star \lesssim 500$ which has already completed, the universe must live at least another $\gtrsim 14$~Gyr. As to humanity we can offer no such guarantees. 

\section{Conclusion}

In this work we investigated constraints on late-time FOPTs, and found that probing the fluctuations provides far stronger constraints than considering only the background evolution. 
This work is one step toward a map of the constraints on FOPTs over all of cosmic time, which limits the most violent dark sector behaviors. Currently the constraints between recombination and dark energy domination are unknown, and remain an obvious target for future exploration. It would also be interesting to consider phase transitions in sectors which are not entirely secluded but have some portal interaction with us. Overall, the signatures of phase transitions in the late universe may be subtle enough that we will not see them if we do not look for them. Since this phenomenon is motivated both on general grounds and in specific theoretical frameworks, it would serve us well to learn how to characterize and search for signals from these drastic rearrangements of the degrees of freedom of the universe.

\section*{Acknowledgements}

We are grateful to Salvatore Bottaro, Raymond Co, Peter Garnavich, Michael Geller, Roni Harnik for helpful discussions. We especially thank Sai Chaitanya Tadepalli for guiding us through the calculation of the two-point correlation function of the composite operators, and Soubhik Kumar for valuable feedback on the draft. This work was partially supported by the NSF Grant PHY-2412701. SK and YT would also like to thank Munich Institute for Astro-, Particle and BioPhysics (MIAPbP) which is funded by the Deutsche Forschungsgemeinschaft (DFG, German Research Foundation) under Germany’s Excellence Strategy – EXC-2094 – 390783311. YT would like to thank the Tom and Carolyn Marquez Chair Fund for its generous support and the Aspen Center for Physics (supported by NSF grant PHY-2210452). YT would also like to thank the host of Fermilab, supported by the URA-22-F-13 fund.

\bibliographystyle{utphys3}
\bibliography{arXiv_v1/latePT} 

\providecommand{\href}[2]{#2}\begingroup\raggedright\begin{thebibliography}{10}

\bibitem{Kosowsky:1992rz}
A.~Kosowsky, M.~S. Turner, and R.~Watkins, ``{Gravitational waves from first
  order cosmological phase transitions},''
  \href{https://dx.doi.org/10.1103/PhysRevLett.69.2026}{{\em Phys. Rev. Lett.}
  {\bfseries 69} (1992) 2026--2029}.

\bibitem{Kosowsky:1991ua}
A.~Kosowsky, M.~S. Turner, and R.~Watkins, ``{Gravitational radiation from
  colliding vacuum bubbles},''
  \href{https://dx.doi.org/10.1103/PhysRevD.45.4514}{{\em Phys. Rev. D}
  {\bfseries 45} (1992) 4514--4535}.

\bibitem{Kosowsky:1992vn}
A.~Kosowsky and M.~S. Turner, ``{Gravitational radiation from colliding vacuum
  bubbles: envelope approximation to many bubble collisions},''
  \href{https://dx.doi.org/10.1103/PhysRevD.47.4372}{{\em Phys. Rev. D}
  {\bfseries 47} (1993) 4372--4391},
  \href{https://arxiv.org/abs/astro-ph/9211004}{{\ttfamily
  arXiv:astro-ph/9211004}}.

\bibitem{Kamionkowski:1993fg}
M.~Kamionkowski, A.~Kosowsky, and M.~S. Turner, ``{Gravitational radiation from
  first order phase transitions},''
  \href{https://dx.doi.org/10.1103/PhysRevD.49.2837}{{\em Phys. Rev. D}
  {\bfseries 49} (1994) 2837--2851},
  \href{https://arxiv.org/abs/astro-ph/9310044}{{\ttfamily
  arXiv:astro-ph/9310044}}.

\bibitem{Caprini:2015zlo}
C.~Caprini {\em et~al.}, ``{Science with the space-based interferometer eLISA.
  II: Gravitational waves from cosmological phase transitions},''
  \href{https://dx.doi.org/10.1088/1475-7516/2016/04/001}{{\em JCAP} {\bfseries
  04} (2016) 001}, \href{https://arxiv.org/abs/1512.06239}{{\ttfamily
  arXiv:1512.06239 [astro-ph.CO]}}.

\bibitem{Caprini:2019egz}
C.~Caprini {\em et~al.}, ``{Detecting gravitational waves from cosmological
  phase transitions with LISA: an update},''
  \href{https://dx.doi.org/10.1088/1475-7516/2020/03/024}{{\em JCAP} {\bfseries
  03} (2020) 024}, \href{https://arxiv.org/abs/1910.13125}{{\ttfamily
  arXiv:1910.13125 [astro-ph.CO]}}.

\bibitem{Caldwell:2022qsj}
R.~Caldwell {\em et~al.}, ``{Detection of early-universe gravitational-wave
  signatures and fundamental physics},''
  \href{https://dx.doi.org/10.1007/s10714-022-03027-x}{{\em Gen. Rel. Grav.}
  {\bfseries 54} no.~12, (2022) 156},
  \href{https://arxiv.org/abs/2203.07972}{{\ttfamily arXiv:2203.07972
  [gr-qc]}}.

\bibitem{LISA:2022kgy}
{\bfseries LISA} Collaboration, K.~G. Arun {\em et~al.}, ``{New horizons for
  fundamental physics with LISA},''
  \href{https://dx.doi.org/10.1007/s41114-022-00036-9}{{\em Living Rev. Rel.}
  {\bfseries 25} no.~1, (2022) 4},
  \href{https://arxiv.org/abs/2205.01597}{{\ttfamily arXiv:2205.01597
  [gr-qc]}}.

\bibitem{LISACosmologyWorkingGroup:2022jok}
{\bfseries LISA Cosmology Working Group} Collaboration, P.~Auclair {\em
  et~al.}, ``{Cosmology with the Laser Interferometer Space Antenna},''
  \href{https://dx.doi.org/10.1007/s41114-023-00045-2}{{\em Living Rev. Rel.}
  {\bfseries 26} no.~1, (2023) 5},
  \href{https://arxiv.org/abs/2204.05434}{{\ttfamily arXiv:2204.05434
  [astro-ph.CO]}}.

\bibitem{Geller:2018mwu}
M.~Geller, A.~Hook, R.~Sundrum, and Y.~Tsai, ``{Primordial Anisotropies in the
  Gravitational Wave Background from Cosmological Phase Transitions},''
  \href{https://dx.doi.org/10.1103/PhysRevLett.121.201303}{{\em Phys. Rev.
  Lett.} {\bfseries 121} no.~20, (2018) 201303},
  \href{https://arxiv.org/abs/1803.10780}{{\ttfamily arXiv:1803.10780
  [hep-ph]}}.

\bibitem{Kumar:2021ffi}
S.~Kumar, R.~Sundrum, and Y.~Tsai, ``{Non-Gaussian stochastic gravitational
  waves from phase transitions},''
  \href{https://dx.doi.org/10.1007/JHEP11(2021)107}{{\em JHEP} {\bfseries 11}
  (2021) 107}, \href{https://arxiv.org/abs/2102.05665}{{\ttfamily
  arXiv:2102.05665 [astro-ph.CO]}}.

\bibitem{LISACosmologyWorkingGroup:2022kbp}
{\bfseries LISA Cosmology Working Group} Collaboration, N.~Bartolo {\em
  et~al.}, ``{Probing anisotropies of the Stochastic Gravitational Wave
  Background with LISA},''
  \href{https://dx.doi.org/10.1088/1475-7516/2022/11/009}{{\em JCAP} {\bfseries
  11} (2022) 009}, \href{https://arxiv.org/abs/2201.08782}{{\ttfamily
  arXiv:2201.08782 [astro-ph.CO]}}.

\bibitem{Bodas:2022urf}
A.~Bodas and R.~Sundrum, ``{Large primordial fluctuations in gravitational
  waves from phase transitions},''
  \href{https://dx.doi.org/10.1007/JHEP06(2023)029}{{\em JHEP} {\bfseries 06}
  (2023) 029}, \href{https://arxiv.org/abs/2211.09301}{{\ttfamily
  arXiv:2211.09301 [hep-ph]}}.

\bibitem{Cui:2023dlo}
Y.~Cui, S.~Kumar, R.~Sundrum, and Y.~Tsai, ``{Unraveling cosmological
  anisotropies within stochastic gravitational wave backgrounds},''
  \href{https://dx.doi.org/10.1088/1475-7516/2023/10/064}{{\em JCAP} {\bfseries
  10} (2023) 064}, \href{https://arxiv.org/abs/2307.10360}{{\ttfamily
  arXiv:2307.10360 [astro-ph.CO]}}.

\bibitem{NANOGrav:2020bcs}
{\bfseries NANOGrav} Collaboration, Z.~Arzoumanian {\em et~al.}, ``{The
  NANOGrav 12.5 yr Data Set: Search for an Isotropic Stochastic
  Gravitational-wave Background},''
  \href{https://dx.doi.org/10.3847/2041-8213/abd401}{{\em Astrophys. J. Lett.}
  {\bfseries 905} no.~2, (2020) L34},
  \href{https://arxiv.org/abs/2009.04496}{{\ttfamily arXiv:2009.04496
  [astro-ph.HE]}}.

\bibitem{NANOGrav:2023gor}
{\bfseries NANOGrav} Collaboration, G.~Agazie {\em et~al.}, ``{The NANOGrav 15
  yr Data Set: Evidence for a Gravitational-wave Background},''
  \href{https://dx.doi.org/10.3847/2041-8213/acdac6}{{\em Astrophys. J. Lett.}
  {\bfseries 951} no.~1, (2023) L8},
  \href{https://arxiv.org/abs/2306.16213}{{\ttfamily arXiv:2306.16213
  [astro-ph.HE]}}.

\bibitem{EPTA:2021crs}
{\bfseries EPTA} Collaboration, S.~Chen {\em et~al.}, ``{Common-red-signal
  analysis with 24-yr high-precision timing of the European Pulsar Timing
  Array: inferences in the stochastic gravitational-wave background search},''
  \href{https://dx.doi.org/10.1093/mnras/stab2833}{{\em Mon. Not. Roy. Astron.
  Soc.} {\bfseries 508} no.~4, (2021) 4970--4993},
  \href{https://arxiv.org/abs/2110.13184}{{\ttfamily arXiv:2110.13184
  [astro-ph.HE]}}.

\bibitem{EPTA:2023fyk}
{\bfseries EPTA, InPTA:} Collaboration, J.~Antoniadis {\em et~al.}, ``{The
  second data release from the European Pulsar Timing Array - III. Search for
  gravitational wave signals},''
  \href{https://dx.doi.org/10.1051/0004-6361/202346844}{{\em Astron.
  Astrophys.} {\bfseries 678} (2023) A50},
  \href{https://arxiv.org/abs/2306.16214}{{\ttfamily arXiv:2306.16214
  [astro-ph.HE]}}.

\bibitem{EPTA:2023xxk}
{\bfseries EPTA, InPTA} Collaboration, J.~Antoniadis {\em et~al.}, ``{The
  second data release from the European Pulsar Timing Array - IV. Implications
  for massive black holes, dark matter, and the early Universe},''
  \href{https://dx.doi.org/10.1051/0004-6361/202347433}{{\em Astron.
  Astrophys.} {\bfseries 685} (2024) A94},
  \href{https://arxiv.org/abs/2306.16227}{{\ttfamily arXiv:2306.16227
  [astro-ph.CO]}}.

\bibitem{Goncharov:2021oub}
B.~Goncharov {\em et~al.}, ``{On the Evidence for a Common-spectrum Process in
  the Search for the Nanohertz Gravitational-wave Background with the Parkes
  Pulsar Timing Array},''
  \href{https://dx.doi.org/10.3847/2041-8213/ac17f4}{{\em Astrophys. J. Lett.}
  {\bfseries 917} no.~2, (2021) L19},
  \href{https://arxiv.org/abs/2107.12112}{{\ttfamily arXiv:2107.12112
  [astro-ph.HE]}}.

\bibitem{Reardon:2023gzh}
D.~J. Reardon {\em et~al.}, ``{Search for an Isotropic Gravitational-wave
  Background with the Parkes Pulsar Timing Array},''
  \href{https://dx.doi.org/10.3847/2041-8213/acdd02}{{\em Astrophys. J. Lett.}
  {\bfseries 951} no.~1, (2023) L6},
  \href{https://arxiv.org/abs/2306.16215}{{\ttfamily arXiv:2306.16215
  [astro-ph.HE]}}.

\bibitem{Xu:2023wog}
H.~Xu {\em et~al.}, ``{Searching for the Nano-Hertz Stochastic Gravitational
  Wave Background with the Chinese Pulsar Timing Array Data Release I},''
  \href{https://dx.doi.org/10.1088/1674-4527/acdfa5}{{\em Res. Astron.
  Astrophys.} {\bfseries 23} no.~7, (2023) 075024},
  \href{https://arxiv.org/abs/2306.16216}{{\ttfamily arXiv:2306.16216
  [astro-ph.HE]}}.

\bibitem{Bringmann:2023opz}
T.~Bringmann, P.~F. Depta, T.~Konstandin, K.~Schmidt-Hoberg, and C.~Tasillo,
  ``{Does NANOGrav observe a dark sector phase transition?},''
  \href{https://dx.doi.org/10.1088/1475-7516/2023/11/053}{{\em JCAP} {\bfseries
  11} (2023) 053}, \href{https://arxiv.org/abs/2306.09411}{{\ttfamily
  arXiv:2306.09411 [astro-ph.CO]}}.

\bibitem{Addazi:2023jvg}
A.~Addazi, Y.-F. Cai, A.~Marciano, and L.~Visinelli, ``{Have pulsar timing
  array methods detected a cosmological phase transition?},''
  \href{https://dx.doi.org/10.1103/PhysRevD.109.015028}{{\em Phys. Rev. D}
  {\bfseries 109} no.~1, (2024) 015028},
  \href{https://arxiv.org/abs/2306.17205}{{\ttfamily arXiv:2306.17205
  [astro-ph.CO]}}.

\bibitem{Ghosh:2023aum}
T.~Ghosh, A.~Ghoshal, H.-K. Guo, F.~Hajkarim, S.~F. King, K.~Sinha, X.~Wang,
  and G.~White, ``{Did we hear the sound of the Universe boiling? Analysis
  using the full fluid velocity profiles and NANOGrav 15-year data},''
  \href{https://dx.doi.org/10.1088/1475-7516/2024/05/100}{{\em JCAP} {\bfseries
  05} (2024) 100}, \href{https://arxiv.org/abs/2307.02259}{{\ttfamily
  arXiv:2307.02259 [astro-ph.HE]}}.

\bibitem{Salvio:2023ynn}
A.~Salvio, ``{Supercooling in radiative symmetry breaking: theory extensions,
  gravitational wave detection and primordial black holes},''
  \href{https://dx.doi.org/10.1088/1475-7516/2023/12/046}{{\em JCAP} {\bfseries
  12} (2023) 046}, \href{https://arxiv.org/abs/2307.04694}{{\ttfamily
  arXiv:2307.04694 [hep-ph]}}.

\bibitem{Winkler:2024olr}
M.~W. Winkler and K.~Freese, ``{Origin of the Stochastic Gravitational Wave
  Background: First-Order Phase Transition vs. Black Hole Mergers},''
  \href{https://arxiv.org/abs/2401.13729}{{\ttfamily arXiv:2401.13729
  [astro-ph.CO]}}.

\bibitem{Banik:2025fnc}
A.~Banik, J.~H. Kim, J.~S. Pi, and Y.~Tsai, ``{Echoes of Self-Interacting Dark
  Matter from Binary Black Hole Mergers},''
  \href{https://arxiv.org/abs/2503.08787}{{\ttfamily arXiv:2503.08787
  [astro-ph.CO]}}.

\bibitem{Riess:2021jrx}
A.~G. Riess {\em et~al.}, ``{A Comprehensive Measurement of the Local Value of
  the Hubble Constant with 1 km/s/Mpc Uncertainty from the Hubble Space
  Telescope and the SH0ES Team},''
  \href{https://dx.doi.org/10.3847/2041-8213/ac5c5b}{{\em Astrophys. J. Lett.}
  {\bfseries 934} no.~1, (2022) L7},
  \href{https://arxiv.org/abs/2112.04510}{{\ttfamily arXiv:2112.04510
  [astro-ph.CO]}}.

\bibitem{Schoneberg:2021qvd}
N.~Sch\"oneberg, G.~Franco~Abell\'an, A.~P\'erez~S\'anchez, S.~J. Witte,
  V.~Poulin, and J.~Lesgourgues, ``{The H0 Olympics: A fair ranking of proposed
  models},'' \href{https://dx.doi.org/10.1016/j.physrep.2022.07.001}{{\em Phys.
  Rept.} {\bfseries 984} (2022) 1--55},
  \href{https://arxiv.org/abs/2107.10291}{{\ttfamily arXiv:2107.10291
  [astro-ph.CO]}}.

\bibitem{Niedermann:2019olb}
F.~Niedermann and M.~S. Sloth, ``{New early dark energy},''
  \href{https://dx.doi.org/10.1103/PhysRevD.103.L041303}{{\em Phys. Rev. D}
  {\bfseries 103} no.~4, (2021) L041303},
  \href{https://arxiv.org/abs/1910.10739}{{\ttfamily arXiv:1910.10739
  [astro-ph.CO]}}.

\bibitem{Niedermann:2020dwg}
F.~Niedermann and M.~S. Sloth, ``{Resolving the Hubble tension with new early
  dark energy},'' \href{https://dx.doi.org/10.1103/PhysRevD.102.063527}{{\em
  Phys. Rev. D} {\bfseries 102} no.~6, (2020) 063527},
  \href{https://arxiv.org/abs/2006.06686}{{\ttfamily arXiv:2006.06686
  [astro-ph.CO]}}.

\bibitem{Niedermann:2023ssr}
F.~Niedermann and M.~S. Sloth, ``{New Early Dark Energy as a solution to the
  $H_0$ and $S_8$ tensions},''
  \href{https://arxiv.org/abs/2307.03481}{{\ttfamily arXiv:2307.03481
  [hep-ph]}}.

\bibitem{Garny:2024ums}
M.~Garny, F.~Niedermann, H.~Rubira, and M.~S. Sloth, ``{Hot new early dark
  energy bridging cosmic gaps: Supercooled phase transition reconciles stepped
  dark radiation solutions to the Hubble tension with BBN},''
  \href{https://dx.doi.org/10.1103/PhysRevD.110.023531}{{\em Phys. Rev. D}
  {\bfseries 110} no.~2, (2024) 023531},
  \href{https://arxiv.org/abs/2404.07256}{{\ttfamily arXiv:2404.07256
  [astro-ph.CO]}}.

\bibitem{Chatrchyan:2024xjj}
A.~Chatrchyan, F.~Niedermann, V.~Poulin, and M.~S. Sloth, ``{Confronting cold
  new early dark energy and its equation of state with updated CMB, supernovae,
  and BAO data},'' \href{https://dx.doi.org/10.1103/PhysRevD.111.043536}{{\em
  Phys. Rev. D} {\bfseries 111} no.~4, (2025) 043536},
  \href{https://arxiv.org/abs/2408.14537}{{\ttfamily arXiv:2408.14537
  [astro-ph.CO]}}.

\bibitem{Greene:2024xgq}
K.~Greene, A.~Ireland, G.~Krnjaic, and Y.~Tsai, ``{Observable CMB B-modes from
  Cosmological Phase Transitions},''
  \href{https://arxiv.org/abs/2410.23348}{{\ttfamily arXiv:2410.23348
  [astro-ph.CO]}}.

\bibitem{Bai:2021ibt}
Y.~Bai and M.~Korwar, ``{Cosmological constraints on first-order phase
  transitions},'' \href{https://dx.doi.org/10.1103/PhysRevD.105.095015}{{\em
  Phys. Rev. D} {\bfseries 105} no.~9, (2022) 095015},
  \href{https://arxiv.org/abs/2109.14765}{{\ttfamily arXiv:2109.14765
  [hep-ph]}}.

\bibitem{Caprini:2007xq}
C.~Caprini, R.~Durrer, and G.~Servant, ``{Gravitational wave generation from
  bubble collisions in first-order phase transitions: An analytic approach},''
  \href{https://dx.doi.org/10.1103/PhysRevD.77.124015}{{\em Phys. Rev. D}
  {\bfseries 77} (2008) 124015},
  \href{https://arxiv.org/abs/0711.2593}{{\ttfamily arXiv:0711.2593
  [astro-ph]}}.

\bibitem{Jinno:2016vai}
R.~Jinno and M.~Takimoto, ``{Gravitational waves from bubble collisions: An
  analytic derivation},''
  \href{https://dx.doi.org/10.1103/PhysRevD.95.024009}{{\em Phys. Rev. D}
  {\bfseries 95} no.~2, (2017) 024009},
  \href{https://arxiv.org/abs/1605.01403}{{\ttfamily arXiv:1605.01403
  [astro-ph.CO]}}.

\bibitem{Jinno:2017fby}
R.~Jinno and M.~Takimoto, ``{Gravitational waves from bubble dynamics: Beyond the Envelope},''
  \href{https://doi.org/10.1088/1475-7516/2019/01/060}{{\em JCAP}
  {\bfseries 01} (2019) 060},
  \href{https://arxiv.org/abs/1707.03111}{{\ttfamily arXiv:1707.03111
  [hep-ph]}}.

\bibitem{Buckley:2024nen}
M.~R.~Buckley, P.~Du, N.~Fernandez, and M.~J.~Weikert,
  ``{Dark radiation isocurvature from cosmological phase transitions},''
  \href{https://doi.org/10.1088/1475-7516/2024/07/031}{{\em JCAP} {\bfseries
  07} (2024) 031}, \href{https://arxiv.org/abs/2402.13309 }{{\ttfamily
  arXiv:2402.13309 [hep-ph]}}.

\bibitem{Liu:2022lvz}
J.~Liu, L.~Bian, R.-G. Cai, Z.-K. Guo, and S.-J. Wang, ``{Constraining
  First-Order Phase Transitions with Curvature Perturbations},''
  \href{https://dx.doi.org/10.1103/PhysRevLett.130.051001}{{\em Phys. Rev.
  Lett.} {\bfseries 130} no.~5, (2023) 051001},
  \href{https://arxiv.org/abs/2208.14086}{{\ttfamily arXiv:2208.14086
  [astro-ph.CO]}}.

\bibitem{Elor:2023xbz}
G.~Elor, R.~Jinno, S.~Kumar, R.~McGehee, and Y.~Tsai, ``{Finite Bubble
  Statistics Constrain Late Cosmological Phase Transitions},''
  \href{https://dx.doi.org/10.1103/PhysRevLett.133.211003}{{\em Phys. Rev.
  Lett.} {\bfseries 133} no.~21, (2024) 211003},
  \href{https://arxiv.org/abs/2311.16222}{{\ttfamily arXiv:2311.16222
  [hep-ph]}}.

\bibitem{Lewicki:2024ghw}
M.~Lewicki, P.~Toczek, and V.~Vaskonen, ``{Black Holes and Gravitational Waves
  from Slow First-Order Phase Transitions},''
  \href{https://dx.doi.org/10.1103/PhysRevLett.133.221003}{{\em Phys. Rev.
  Lett.} {\bfseries 133} no.~22, (2024) 221003},
  \href{https://arxiv.org/abs/2402.04158}{{\ttfamily arXiv:2402.04158
  [astro-ph.CO]}}.

\bibitem{Zou:2025sow}
J.~Zou, Z.~Zhu, Z.~Zhao, and L.~Bian, ``{Numerical simulations of density
  perturbation and gravitational wave production from cosmological first-order
  phase transition},'' \href{https://arxiv.org/abs/2502.20166}{{\ttfamily
  arXiv:2502.20166 [hep-ph]}}.

\bibitem{Franciolini:2025ztf}
G.~Franciolini, Y.~Gouttenoire, and R.~Jinno, ``{Curvature Perturbations from
  First-Order Phase Transitions: Implications to Black Holes and Gravitational
  Waves},'' \href{https://arxiv.org/abs/2503.01962}{{\ttfamily arXiv:2503.01962
  [hep-ph]}}.

\bibitem{Ramberg:2022irf}
N.~Ramberg, W.~Ratzinger, and P.~Schwaller, ``{One {\ensuremath{\mu}} to rule
  them all: CMB spectral distortions can probe domain walls, cosmic strings and
  low scale phase transitions},''
  \href{https://dx.doi.org/10.1088/1475-7516/2023/02/039}{{\em JCAP} {\bfseries
  02} (2023) 039}, \href{https://arxiv.org/abs/2209.14313}{{\ttfamily
  arXiv:2209.14313 [hep-ph]}}.

\bibitem{Geller}
S.~Bottaro, M.~Geller, D.~Redigolo, and M.~Tsur {\em in preparation} .

\bibitem{Abbott:1984qf}
L.~F. Abbott, ``{A Mechanism for Reducing the Value of the Cosmological
  Constant},'' \href{https://dx.doi.org/10.1016/0370-2693(85)90459-9}{{\em
  Phys. Lett. B} {\bfseries 150} (1985) 427--430}.

\bibitem{Brown:1987dd}
J.~D. Brown and C.~Teitelboim, ``{Dynamical Neutralization of the Cosmological
  Constant},'' \href{https://dx.doi.org/10.1016/0370-2693(87)91190-7}{{\em
  Phys. Lett. B} {\bfseries 195} (1987) 177--182}.

\bibitem{Brown:1988kg}
J.~D. Brown and C.~Teitelboim, ``{Neutralization of the Cosmological Constant
  by Membrane Creation},''
  \href{https://dx.doi.org/10.1016/0550-3213(88)90559-7}{{\em Nucl. Phys. B}
  {\bfseries 297} (1988) 787--836}.

\bibitem{Duncan:1989ug}
M.~J. Duncan and L.~G. Jensen, ``{Four Forms and the Vanishing of the
  Cosmological Constant},''
  \href{https://dx.doi.org/10.1016/0550-3213(90)90344-D}{{\em Nucl. Phys. B}
  {\bfseries 336} (1990) 100--114}.

\bibitem{Bousso:2000xa}
R.~Bousso and J.~Polchinski, ``{Quantization of four form fluxes and dynamical
  neutralization of the cosmological constant},''
  \href{https://dx.doi.org/10.1088/1126-6708/2000/06/006}{{\em JHEP} {\bfseries
  06} (2000) 006}, \href{https://arxiv.org/abs/hep-th/0004134}{{\ttfamily
  arXiv:hep-th/0004134}}.

\bibitem{Feng:2000if}
J.~L. Feng, J.~March-Russell, S.~Sethi, and F.~Wilczek, ``{Saltatory relaxation
  of the cosmological constant},''
  \href{https://dx.doi.org/10.1016/S0550-3213(01)00097-9}{{\em Nucl. Phys. B}
  {\bfseries 602} (2001) 307--328},
  \href{https://arxiv.org/abs/hep-th/0005276}{{\ttfamily
  arXiv:hep-th/0005276}}.

\bibitem{Dvali:2001sm}
G.~R. Dvali and A.~Vilenkin, ``{Field theory models for variable cosmological
  constant},'' \href{https://dx.doi.org/10.1103/PhysRevD.64.063509}{{\em Phys.
  Rev. D} {\bfseries 64} (2001) 063509},
  \href{https://arxiv.org/abs/hep-th/0102142}{{\ttfamily
  arXiv:hep-th/0102142}}.

\bibitem{Kaloper:2022jpv}
N.~Kaloper and A.~Westphal, ``{Quantum-mechanical mechanism for reducing the
  cosmological constant},''
  \href{https://dx.doi.org/10.1103/PhysRevD.106.L101701}{{\em Phys. Rev. D}
  {\bfseries 106} no.~10, (2022) L101701},
  \href{https://arxiv.org/abs/2204.13124}{{\ttfamily arXiv:2204.13124
  [hep-th]}}.

\bibitem{Kaloper:2022yiw}
N.~Kaloper, ``{General relativity on the multiverse and
  nature\textquoteright{}s hierarchies},''
  \href{https://dx.doi.org/10.1103/PhysRevD.106.044023}{{\em Phys. Rev. D}
  {\bfseries 106} no.~4, (2022) 044023},
  \href{https://arxiv.org/abs/2202.08860}{{\ttfamily arXiv:2202.08860
  [hep-th]}}.

\bibitem{Kaloper:2023kua}
N.~Kaloper, ``{Axion flux monodromy discharges relax the cosmological
  constant},'' \href{https://dx.doi.org/10.1088/1475-7516/2023/11/032}{{\em
  JCAP} {\bfseries 11} (2023) 032},
  \href{https://arxiv.org/abs/2307.10365}{{\ttfamily arXiv:2307.10365
  [hep-th]}}.

\bibitem{Kaloper:2023xfl}
N.~Kaloper, ``{de Sitter space decay and cosmological constant relaxation in
  unimodular gravity with charged membranes},''
  \href{https://dx.doi.org/10.1103/PhysRevD.108.025005}{{\em Phys. Rev. D}
  {\bfseries 108} no.~2, (2023) 025005},
  \href{https://arxiv.org/abs/2305.02349}{{\ttfamily arXiv:2305.02349
  [hep-th]}}.

\bibitem{Kaloper:2025goq}
N.~Kaloper, ``{Discretely Evanescent Dark Energy},''
  \href{https://arxiv.org/abs/2506.04317}{{\ttfamily arXiv:2506.04317
  [hep-th]}}.

\bibitem{Dvali:2003br}
G.~Dvali and A.~Vilenkin, ``{Cosmic attractors and gauge hierarchy},''
  \href{https://dx.doi.org/10.1103/PhysRevD.70.063501}{{\em Phys. Rev. D}
  {\bfseries 70} (2004) 063501},
  \href{https://arxiv.org/abs/hep-th/0304043}{{\ttfamily
  arXiv:hep-th/0304043}}.

\bibitem{Dvali:2004tma}
G.~Dvali, ``{Large hierarchies from attractor vacua},''
  \href{https://dx.doi.org/10.1103/PhysRevD.74.025018}{{\em Phys. Rev. D}
  {\bfseries 74} (2006) 025018},
  \href{https://arxiv.org/abs/hep-th/0410286}{{\ttfamily
  arXiv:hep-th/0410286}}.

\bibitem{Graham:2015cka}
P.~W. Graham, D.~E. Kaplan, and S.~Rajendran, ``{Cosmological Relaxation of the
  Electroweak Scale},''
  \href{https://dx.doi.org/10.1103/PhysRevLett.115.221801}{{\em Phys. Rev.
  Lett.} {\bfseries 115} no.~22, (2015) 221801},
  \href{https://arxiv.org/abs/1504.07551}{{\ttfamily arXiv:1504.07551
  [hep-ph]}}.

\bibitem{Kaplan:2015fuy}
D.~E. Kaplan and R.~Rattazzi, ``{Large field excursions and approximate
  discrete symmetries from a clockwork axion},''
  \href{https://dx.doi.org/10.1103/PhysRevD.93.085007}{{\em Phys. Rev. D}
  {\bfseries 93} no.~8, (2016) 085007},
  \href{https://arxiv.org/abs/1511.01827}{{\ttfamily arXiv:1511.01827
  [hep-ph]}}.

\bibitem{Choi:2015fiu}
K.~Choi and S.~H. Im, ``{Realizing the relaxion from multiple axions and its UV
  completion with high scale supersymmetry},''
  \href{https://dx.doi.org/10.1007/JHEP01(2016)149}{{\em JHEP} {\bfseries 01}
  (2016) 149}, \href{https://arxiv.org/abs/1511.00132}{{\ttfamily
  arXiv:1511.00132 [hep-ph]}}.

\bibitem{Ibanez:2015fcv}
L.~E. Ibanez, M.~Montero, A.~Uranga, and I.~Valenzuela, ``{Relaxion Monodromy
  and the Weak Gravity Conjecture},''
  \href{https://dx.doi.org/10.1007/JHEP04(2016)020}{{\em JHEP} {\bfseries 04}
  (2016) 020}, \href{https://arxiv.org/abs/1512.00025}{{\ttfamily
  arXiv:1512.00025 [hep-th]}}.

\bibitem{Gupta:2015uea}
R.~S. Gupta, Z.~Komargodski, G.~Perez, and L.~Ubaldi, ``{Is the Relaxion an
  Axion?},'' \href{https://dx.doi.org/10.1007/JHEP02(2016)166}{{\em JHEP}
  {\bfseries 02} (2016) 166},
  \href{https://arxiv.org/abs/1509.00047}{{\ttfamily arXiv:1509.00047
  [hep-ph]}}.

\bibitem{Hook:2016mqo}
A.~Hook and G.~Marques-Tavares, ``{Relaxation from particle production},''
  \href{https://dx.doi.org/10.1007/JHEP12(2016)101}{{\em JHEP} {\bfseries 12}
  (2016) 101}, \href{https://arxiv.org/abs/1607.01786}{{\ttfamily
  arXiv:1607.01786 [hep-ph]}}.

\bibitem{Banks:1991mb}
T.~Banks, M.~Dine, and N.~Seiberg, ``{Irrational axions as a solution of the
  strong CP problem in an eternal universe},''
  \href{https://dx.doi.org/10.1016/0370-2693(91)90561-4}{{\em Phys. Lett. B}
  {\bfseries 273} (1991) 105--110},
  \href{https://arxiv.org/abs/hep-th/9109040}{{\ttfamily
  arXiv:hep-th/9109040}}.

\bibitem{Dvali:2005zk}
G.~Dvali, ``{A Vacuum accumulation solution to the strong CP problem},''
  \href{https://dx.doi.org/10.1103/PhysRevD.74.025019}{{\em Phys. Rev. D}
  {\bfseries 74} (2006) 025019},
  \href{https://arxiv.org/abs/hep-th/0510053}{{\ttfamily
  arXiv:hep-th/0510053}}.

\bibitem{Kaloper:2017fsa}
N.~Kaloper and J.~Terning, ``{Landscaping the Strong CP Problem},''
  \href{https://dx.doi.org/10.1007/JHEP03(2019)032}{{\em JHEP} {\bfseries 03}
  (2019) 032}, \href{https://arxiv.org/abs/1710.01740}{{\ttfamily
  arXiv:1710.01740 [hep-th]}}.

\bibitem{Kaloper:2025upu}
N.~Kaloper, ``{A Quantal Theory of Restoration of Strong CP Symmetry},''
  \href{https://arxiv.org/abs/2505.04690}{{\ttfamily arXiv:2505.04690
  [hep-ph]}}.

\bibitem{Kaloper:2025wgn}
N.~Kaloper, ``{An Alternative to Axion},''
  \href{https://arxiv.org/abs/2504.21078}{{\ttfamily arXiv:2504.21078
  [hep-ph]}}.

\bibitem{Fujii:1982ms}
Y.~Fujii, ``{Origin of the Gravitational Constant and Particle Masses in Scale
  Invariant Scalar - Tensor Theory},''
  \href{https://dx.doi.org/10.1103/PhysRevD.26.2580}{{\em Phys. Rev. D}
  {\bfseries 26} (1982) 2580}.

\bibitem{Ford:1987de}
L.~H. Ford, ``{COSMOLOGICAL CONSTANT DAMPING BY UNSTABLE SCALAR FIELDS},''
  \href{https://dx.doi.org/10.1103/PhysRevD.35.2339}{{\em Phys. Rev. D}
  {\bfseries 35} (1987) 2339}.

\bibitem{Ratra:1987rm}
B.~Ratra and P.~J.~E. Peebles, ``{Cosmological Consequences of a Rolling
  Homogeneous Scalar Field},''
  \href{https://dx.doi.org/10.1103/PhysRevD.37.3406}{{\em Phys. Rev. D}
  {\bfseries 37} (1988) 3406}.

\bibitem{Wetterich:1987fm}
C.~Wetterich, ``{Cosmology and the Fate of Dilatation Symmetry},''
  \href{https://dx.doi.org/10.1016/0550-3213(88)90193-9}{{\em Nucl. Phys. B}
  {\bfseries 302} (1988) 668--696},
  \href{https://arxiv.org/abs/1711.03844}{{\ttfamily arXiv:1711.03844
  [hep-th]}}.

\bibitem{DESI:2025zgx}
{\bfseries DESI} Collaboration, M.~Abdul~Karim {\em et~al.}, ``{DESI DR2
  Results II: Measurements of Baryon Acoustic Oscillations and Cosmological
  Constraints},'' \href{https://arxiv.org/abs/2503.14738}{{\ttfamily
  arXiv:2503.14738 [astro-ph.CO]}}.

\bibitem{HOGAN1983172}
C.~J. Hogan, ``Nucleation of cosmological phase transitions,''
  \href{https://dx.doi.org/https://doi.org/10.1016/0370-2693(83)90553-1}{{\em
  Physics Letters B} {\bfseries 133} no.~3, (1983) 172--176}.
  \url{https://www.sciencedirect.com/science/article/pii/0370269383905531}.

\bibitem{PhysRevD.45.3415}
K.~Enqvist, J.~Ignatius, K.~Kajantie, and K.~Rummukainen, ``Nucleation and
  bubble growth in a first-order cosmological electroweak phase transition,''
  \href{https://dx.doi.org/10.1103/PhysRevD.45.3415}{{\em Phys. Rev. D}
  {\bfseries 45} (May, 1992) 3415--3428}.
  \url{https://link.aps.org/doi/10.1103/PhysRevD.45.3415}.

\bibitem{Hindmarsh:2015qta}
M.~Hindmarsh, S.~J. Huber, K.~Rummukainen, and D.~J. Weir, ``{Numerical
  simulations of acoustically generated gravitational waves at a first order
  phase transition},''
  \href{https://dx.doi.org/10.1103/PhysRevD.92.123009}{{\em Phys. Rev. D}
  {\bfseries 92} no.~12, (2015) 123009},
  \href{https://arxiv.org/abs/1504.03291}{{\ttfamily arXiv:1504.03291
  [astro-ph.CO]}}.

\bibitem{ATHRON2024104094}
P.~Athron, C.~Balázs, A.~Fowlie, L.~Morris, and L.~Wu, ``Cosmological phase
  transitions: From perturbative particle physics to gravitational waves,''
  \href{https://dx.doi.org/https://doi.org/10.1016/j.ppnp.2023.104094}{{\em
  Progress in Particle and Nuclear Physics} {\bfseries 135} (2024) 104094}.
  \url{https://www.sciencedirect.com/science/article/pii/S0146641023000753}.

\bibitem{Planck:2018vyg}
{\bfseries Planck} Collaboration, N.~Aghanim {\em et~al.}, ``{Planck 2018
  results. VI. Cosmological parameters},''
  \href{https://dx.doi.org/10.1051/0004-6361/201833910}{{\em Astron.
  Astrophys.} {\bfseries 641} (2020) A6},
  \href{https://arxiv.org/abs/1807.06209}{{\ttfamily arXiv:1807.06209
  [astro-ph.CO]}}. [Erratum: Astron.Astrophys. 652, C4 (2021)].

\bibitem{planckarchiv}
\url{https://pla.esac.esa.int/#home}.

\bibitem{Sachs:1967er}
R.~K. Sachs and A.~M. Wolfe, ``{Perturbations of a cosmological model and
  angular variations of the microwave background},''
  \href{https://dx.doi.org/10.1007/s10714-007-0448-9}{{\em Astrophys. J.}
  {\bfseries 147} (1967) 73--90}.

\bibitem{DESI:2025fii}
{\bfseries DESI} Collaboration, K.~Lodha {\em et~al.}, ``{Extended Dark Energy
  analysis using DESI DR2 BAO measurements},''
  \href{https://arxiv.org/abs/2503.14743}{{\ttfamily arXiv:2503.14743
  [astro-ph.CO]}}.

\bibitem{ParticleDataGroup:2024cfk}
{\bfseries Particle Data Group} Collaboration, S.~Navas {\em et~al.}, ``{Review
  of particle physics},''
  \href{https://dx.doi.org/10.1103/PhysRevD.110.030001}{{\em Phys. Rev. D}
  {\bfseries 110} no.~3, (2024) 030001}.

\bibitem{Obied:2018sgi}
G.~Obied, H.~Ooguri, L.~Spodyneiko, and C.~Vafa, ``{De Sitter Space and the
  Swampland},'' \href{https://arxiv.org/abs/1806.08362}{{\ttfamily
  arXiv:1806.08362 [hep-th]}}.

\end{thebibliography}\endgroup

\newpage
\appendix
\section{More details of the power spectrum calculation}\label{app.A}
\begin{figure}
    \centering
    \includegraphics[width=1\linewidth]{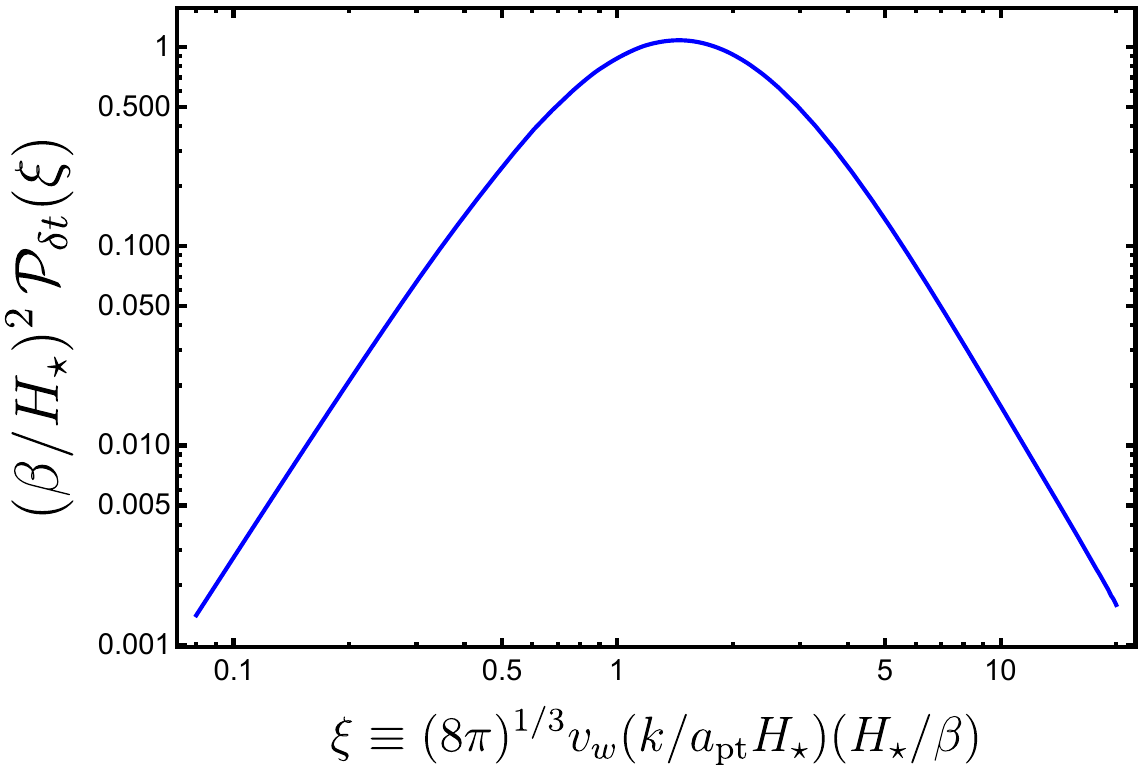}
    \caption{Dimensionless power spectrum of PT time fluctuations from~\cite{Elor:2023xbz}, rescaled by $(\beta/H_\star)^2$, plotted vs. the comoving wavenumber ratio $\xi$, which characterizes the mode relative to the bubble separation.}
    \label{fig:Pdt}
\end{figure}

Fig.~\ref{fig:Pdt} shows the rescaled power spectrum $(\beta/H_\star)^2\mathcal{P}_{\delta t}(\xi)$ of PT time fluctuations, with $\xi$ defined as the comoving wavenumber $k$ times the average bubble size $d_b=(8\pi)^{1/3}v_w/(a_{\rm pt}\beta)$ right before the PT completes. In this parametrization, the curve is insensitive to $(\beta/H_\star, z_{\rm pt})$. We use this curve for the power spectrum in Eq.~(\ref{eq.Pdz0}).

To determine the parametric dependence of the CMB photon redshift perturbation in Eq.~(\ref{eq.dz0}), we use the relation in Eq.~(\ref{eq.chieq}):
\begin{widetext}
\begin{eqnarray}
\int_{0}^{\bar{z}_{\rm pt}}\frac{{\rm d}z}{(1+z)\sqrt{(1-r)\Omega_\Lambda+r\Omega_\Lambda\left(\frac{1+z}{1+\bar{z}_{\rm pt}}\right)^4+\Omega_m(1+z)^3}}&+&\int_{\bar{z}_{\rm pt}}^{\bar{z}_{\rm pt}+\delta z_{\rm pt}}\frac{{\rm d}z}{(1+z)\sqrt{\Omega_\Lambda+\Omega_m(1+z)^3}}\nonumber
\\
=\int_{\delta z_0}^{\bar{z}_{\rm pt}}\frac{{\rm d}z}{(1+z)\sqrt{(1-r)\Omega_\Lambda+r\Omega_\Lambda\left(\frac{1+z}{1+\bar{z}_{\rm pt}}\right)^4+\Omega_m(1+z)^3}}&+&\int_{\bar{z}_{\rm pt}}^{\bar{z}_{\rm pt}+\delta z_{\rm pt}}\frac{{\rm d}z}{(1+z)\sqrt{(1-r)\Omega_\Lambda+r\Omega_\Lambda\left(\frac{1+z}{1+\bar{z}_{\rm pt}+\delta z_{\rm pt}}\right)^4+\Omega_m(1+z)^3}}\,.\nonumber
\end{eqnarray}
\end{widetext}
Our goal is to extract the leading-order $\delta z_{\rm pt}$ dependence in $\delta z_0$ from the equality, using $\delta z_0, \delta z_{\rm pt} \ll r \bar{z}_{\rm pt} \ll 1$.

Combining the first integrals on both sides of the equality, and using $\Omega_\Lambda+\Omega_m \approx 1$ together with the small-redshift approximation, we obtain
\begin{align}
&\int_{0}^{\delta z_{0}}\frac{{\rm d}z}{(1+z)\sqrt{(1-r)\Omega_\Lambda+r\Omega_\Lambda\left(\frac{1+z}{1+\bar{z}_{\rm pt}}\right)^4+\Omega_m(1+z)^3}}\nonumber\\
&\approx \delta z_0\,.\label{eq.approx1}
\end{align}
For the second integral on the right-hand side, the denominator can be expanded as
\begin{eqnarray}
&\approx&\int_{\bar{z}_{\rm pt}}^{\bar{z}_{\rm pt}+\delta z_{\rm pt}}\frac{{\rm d}z\left[1-\frac{2\,r\,\Omega_\Lambda(z-\bar{z}_{\rm pt}-\delta z_{\rm pt})}{\Omega_\Lambda+\Omega_m(1+z)^3}\right]}{(1+z)\sqrt{\Omega_\Lambda+\Omega_m(1+z)^3}}\,.
\end{eqnarray}
Here we use the fact that $r\bar{z}_{\rm pt}\ll 1$ for the PT we consider. Combining this with the second integral on the left-hand side, we approximate the remainder as
\begin{equation}\label{eq.approx2}
\approx \frac{\delta z_{\rm pt}^2\,r\,\Omega_\Lambda}{(1+\bar{z}_{\rm pt})\left[\Omega_{\Lambda}+\Omega_m(1+\bar{z}_{\rm pt})^3\right]^{3/2}}\,.
\end{equation}
Eqs.~(\ref{eq.approx1}) and (\ref{eq.approx2}) give the expression 
\begin{equation}
\delta z_0\approx \delta z_{\rm pt}^2\left[\frac{r\,\Omega_\Lambda}{(1+\bar{z}_{\rm pt})\left[\Omega_\Lambda+\Omega_{\rm m}(1+\bar{z}_{\rm pt})^3\right]^{3/2}}\right]\,
\end{equation}
in Eq.~(\ref{eq.dz0}). The approximation agrees with the full numerical integral to within the percent level.

To obtain the power spectrum $\mathcal{P}_{\delta z_0}(k)$, we need to calculate the two-point correlation function
\begin{equation}\label{eq.z0xz0y}
\langle\delta z_0(x)\delta z_0(y)\rangle=\int\frac{{\rm d}^3k}{(2\pi)^3}e^{i\vec{k}.\left(\vec{x}-\vec{y}\right)}\frac{2\pi^2}{k^3}\mathcal{P}_{\delta z_0}(k)\,.
\end{equation}
From the discussion above, the two-point correlation function comes from the correlation of two composite operators
\begin{align}
\langle\delta z_0(x)\delta z_0(y)\rangle\approx &(1+\bar{z}_{\rm pt})^{-2}r^2\,\Omega_\Lambda^2[\Omega_\Lambda+\Omega_{\rm m}(1+\bar{z}_{\rm pt})^3]^{-3}\nonumber
\\
&\langle\delta z_{\rm pt}^2(x)\delta z_{\rm pt}^2(y)\rangle\,,\label{eq.z0xz0y2}
\end{align}
and we want to express $\langle\delta z_{\rm pt}^2(x)\delta z_{\rm pt}^2(y)\rangle$ in terms of $\mathcal{P}_{\delta t}(k)$. We first define the Fourier transform\footnote{We thank Sai Chaitanya Tadepalli for providing the following calculation of the two-point correlation function of the composite operators.}
\begin{align}
&F[\delta z_{\rm pt}^{2}(x)] =\int d^{3}xe^{-i\vec{k}.\vec{x}}\delta z_{\rm pt}^{2}(x)\,,\\
&=\int d^{3}xe^{-i\vec{k}.\vec{x}}\left[\int\frac{d^{3}p}{\left(2\pi\right)^{3}}e^{i\vec{p}.\vec{x}}\delta z_{{\rm pt}}(\vec{p})\int\frac{d^{3}q}{\left(2\pi\right)^{3}}e^{i\vec{q}.\vec{x}}\delta z_{{\rm pt}}(\vec{q})\right],\nonumber\\
 & =\int\frac{d^{3}p}{\left(2\pi\right)^{3}}\int\frac{d^{3}q}{\left(2\pi\right)^{3}}\left(\int d^{3}xe^{i\left(\vec{p}+\vec{q}-\vec{k}\right).\vec{x}}\right)\delta z_{{\rm pt}}(\vec{p})\delta z_{{\rm pt}}(\vec{q})\,.\nonumber
\end{align}
Integrate over $\vec{x}$ and $\vec{q}$ gives:
\begin{align}
\delta z_{{\rm pt}}^{2}(\vec{k}) & =\int\frac{d^{3}p}{\left(2\pi\right)^{3}}\delta z_{\rm pt}(\vec{p})\delta z_{\rm pt}(\vec{k}-\vec{p})\,.
\end{align}
The correlation function becomes
\begin{align}\label{eq.dz2dz2}
&\langle\delta z_{\rm pt}^2(x)\delta z_{\rm pt}^2(y)\rangle
\\ 
&=\int\frac{d^{3}k}{\left(2\pi\right)^{3}}e^{i\vec{k}.\vec{x}}\int\frac{d^{3}p}{\left(2\pi\right)^{3}}e^{i\vec{p}.\vec{y}}\left\langle \delta z^2_{\rm pt}(\vec{k})\delta z^2_{\rm pt}(\vec{p})\right\rangle \nonumber\\
&=\int\frac{d^{3}k}{\left(2\pi\right)^{3}}\int\frac{d^{3}p}{\left(2\pi\right)^{3}}\int\frac{d^{3}r}{\left(2\pi\right)^{3}}\int\frac{d^{3}q}{\left(2\pi\right)^{3}}e^{i\vec{k}.\vec{x}}e^{i\vec{p}.\vec{y}}\nonumber \\
&\qquad\qquad\left\langle \delta z_{\rm pt}(\vec{r})\delta z_{\rm pt}(\vec{k}-\vec{r})\delta z_{\rm pt}(\vec{q})\delta z_{\rm pt}(\vec{p}-\vec{q})\right\rangle\,, \nonumber
\end{align}
where $\langle...\rangle$ is the equal-time correlator of perturbations in momentum space. 
Only the two-point correlations between perturbations at $x$- and $y$ contribute to the scale-dependent power spectrum for the cosmological measurement:
\begin{eqnarray}
&\left\langle \delta z_{\rm pt}(\vec{r})\delta z_{\rm pt}(\vec{k}-\vec{r})\delta z_{\rm pt}(\vec{q})\delta z_{\rm pt}(\vec{p}-\vec{q})\right\rangle \\
&=\left\langle \delta z_{\rm pt}(\vec{r})\delta z_{\rm pt}(\vec{q})\right\rangle \left\langle \delta z_{\rm pt}(\vec{k}-\vec{r})\delta z_{\rm pt}(\vec{p}-\vec{q})\right\rangle, \nonumber\\
&+\left\langle \delta z_{\rm pt}(\vec{r})\delta z_{\rm pt}(\vec{p}-\vec{q})\right\rangle \left\langle \delta z_{\rm pt}(\vec{k}-\vec{r})\delta z_{\rm pt}(\vec{q})\right\rangle.\nonumber
\end{eqnarray}
Using the definition of the dimensionless power spectrum
\begin{equation}
\left\langle \delta z_{\rm pt}(\vec{k})\delta z_{\rm pt}(\vec{p})\right\rangle =\left(2\pi\right)^{3}\delta(\vec{k}+\vec{p})\frac{2\pi^2}{k^{3}}\mathcal{P}_{\delta z_{\rm pt}}(k)\,,
\end{equation}
we perform the volume integrals in Eq.~(\ref{eq.dz2dz2}) over $\vec{q}$ and $\vec{p}$, which correspond to the connected diagrams, and obtain
\begin{align}\label{eq.dz2result}
&\langle\delta z_{\rm pt}^2(x)\delta z_{\rm pt}^2(y)\rangle_{{\rm conn}} \\
& =\int\frac{d^{3}k}{\left(2\pi\right)^{3}}e^{i\vec{k}.\left(\vec{x}_{1}-\vec{x}_{2}\right)}\int\frac{d^{3}r}{\left(2\pi\right)^{3}}\frac{8\pi^4}{r^{3}|\vec{k}-\vec{r}|^{3}}\mathcal{P}_{\delta z_{\rm pt}}(r)\mathcal{P}_{\delta z_{\rm pt}}(|\vec{k}-\vec{r}|),\nonumber \\
& =\int\frac{d^{3}k}{\left(2\pi\right)^{3}}e^{i\vec{k}.\left(\vec{x}_{1}-\vec{x}_{2}\right)}\left(\int\frac{d r}{r}2\pi^2\mathcal{P}_{\delta z_{\rm pt}}(r)\int_{-1}^1{\rm d}\mu \,s^{-3}\mathcal{P}_{\delta z_{\rm pt}}(s)\right)\,,\nonumber
\end{align}
where $s=\sqrt{k^2+r^2-2kr\mu}$. With Eqs.~(\ref{eq.z0xz0y}), (\ref{eq.z0xz0y2}), (\ref{eq.dz2result}) and 
\begin{equation}
\mathcal{P}_{\delta z_{\rm pt}}(k)\approx(1+\bar{z}_{\rm pt})^2{\cal P}_{\delta t}(k)\,,
\end{equation}
which follows from $\delta a_{\rm pt}/\bar{a}_{\rm pt}=-\delta z_{\rm pt}/(1+\bar{z}_{\rm pt})=H_\star \delta t$, we obtain
\begin{align}
\mathcal{P}_{\delta z_0}(k)\approx&\,\Omega_\Lambda^2[\Omega_\Lambda+\Omega_{\rm m}(1+\bar{z}_{\rm pt})^3]^{-3}\\
&k^3\int\frac{d r}{r}\mathcal{P}_{\delta z_{\rm pt}}(r)\int_{-1}^1{\rm d}\mu \,s^{-3}\mathcal{P}_{\delta z_{\rm pt}}(s)\,.
\end{align}
This is the power spectrum we use for the CMB calculation in Eq.~(\ref{eq.Pdz0}).

\end{document}